\documentclass[letterpaper]{article} 
\usepackage{aaai2026}
\usepackage{times}  
\usepackage{helvet}  
\usepackage{courier}  
\usepackage[hyphens]{url}  
\usepackage{graphicx} 
\usepackage{natbib}  
\usepackage{caption} 
\usepackage{algorithm}
\usepackage{algorithmic}
\usepackage{soul}
\usepackage[utf8]{inputenc}
\usepackage{amsmath}
\usepackage{amsthm}
\usepackage{booktabs}
\usepackage{multirow}
\usepackage{xcolor}
\usepackage{float}
\usepackage{comment}
\usepackage{newfloat}
\usepackage{listings}
\DeclareCaptionStyle{ruled}{labelfont=normalfont,labelsep=colon,strut=off} 
\floatstyle{ruled}
\newfloat{listing}{tb}{lst}{}
\floatname{listing}{Listing}
\title{Dopamine Audiobook: A Training-free MLLM Agent for Emotional and Immersive Audiobook Generation}
\author{
    Yan Rong\textsuperscript{\rm 1}, 
    Shan Yang\textsuperscript{\rm 2}, 
    Chenxing Li\textsuperscript{\rm 2}, 
    Dong Yu\textsuperscript{\rm 2}, 
    Li Liu\textsuperscript{\rm 1}\thanks{Corresponding Author: avrillliu@hkust-gz.edu.cn}
}
\affiliations{
     \textsuperscript{\rm 1}The Hong Kong University of Science and Technology (Guangzhou)\\
     \textsuperscript{\rm 2}Tencent AI Lab\\
}

\usepackage{bibentry}

\begin{document}

\maketitle
\begin{abstract}
Audiobook generation aims to create rich, immersive listening experiences from multimodal inputs, but current approaches face three critical challenges: (1) the lack of synergistic generation of diverse audio types (\textit{e.g.,} speech, sound effects, and music) with precise temporal and semantic alignment; (2) the difficulty in conveying expressive, fine-grained emotions, which often results in machine-like vocal outputs; and (3) the absence of automated evaluation frameworks that align with human preferences for complex and diverse audio.
To address these issues, we propose \textbf{Dopamine Audiobook}, a novel unified training-free multi-agent system, where a multimodal large language model (MLLM) serves two specialized roles (\textit{i.e.}, speech designer and audio designer) for emotional, human-like, and immersive audiobook generation and evaluation.
Specifically, we firstly propose a flow-based, context-aware framework for diverse audio generation with word-level semantic and temporal alignment. 
To enhance expressiveness, we then design word-level paralinguistic augmentation, utterance-level prosody retrieval, and adaptive TTS model selection. 
Finally, for evaluation, we introduce a novel MLLM-based evaluation framework incorporating self-critique, perspective-taking, and psychological MagicEmo prompts to ensure human-aligned and self-aligned assessments.
Experimental results demonstrate that our method achieves state-of-the-art (SOTA) performance on multiple metrics. Importantly, our evaluation framework shows better alignment with human preferences and transferability across audio tasks.
\end{abstract}

\begin{links}
    \link{Demo Page}{https://dopamine-audiobook.github.io}
\end{links}

\section{Introduction}
Audiobook generation~\cite{de2024personalized,tan2024audioxtend} refers to the process of producing rich and engaging narrated audio works that align with multi-modal input instructions. The ultimate goal is to create truly immersive listening experiences by synergistically combining expressive speech with contextual sound effects and fitting background music. This technology opens up new opportunities for diverse multimodal human-computer interactions, such as barrier-free reading~\cite{nadeau2024reading}, interactive VR/AR experiences~\cite{dam2024taxonomy}, and personalized content creation~\cite{rong2025seeing}.

In the literature, most efforts have primarily focused on text-to-speech (TTS) generation~\cite{chen2024f5,du2024cosyvoice2,xie2024towards}, which limits audiobooks to only speech or dialogue portions and often lacks a true sense of immersion. Crafting an emotional, human-like, and truly immersive audio production that integrates diverse audio types (\textit{e.g.}, speech, sound effect, and music) presents \textbf{three main challenges}:
\textbf{(1)} The synergistic generation of speech, sound effects (SFX), and background music (BGM) with perfect temporal and semantic alignment is crucial for creating immersive audio experiences. However, current approaches that rely on separate models for each component~\cite{cheng2025mmaudio,zhang2025inspiremusic} often produce disjointed and unnatural-sounding outputs. Achieving coherent planning and precise time-aligned fusion, where SFX are triggered at specific word mentions and BGM dynamically adapts to narrative shifts, remains a complex, largely unsolved coordination problem.
\textbf{(2)} Current TTS methods struggle to generate dialogue with rich and fine-grained emotions, often producing an ``\textbf{averaged emotion}" that fails to capture the dynamic shifts in a narrative. Besides, the generated speech often lacks paralinguistic features, resulting in machine-like outputs. Individual TTS models also exhibit limited scene adaptability, excelling only in specific scenarios and failing to meet diverse user demands.
\textbf{(3)} There is still a lack of an automated human-aligned evaluation framework. Existing methods are either automated but misaligned with human preferences or rely on costly, time-consuming human evaluations.

Therefore, in this work, the main question we aim to address is: \textit{How to develop an integrated framework that generates emotional and immersive audiobooks, complete with an automatic human-aligned evaluation framework that effectively satisfies various scenarios?}

To address the above challenges, we propose \textbf{Dopamine Audiobook}, a novel unified training-free multi-agent system (MAS), where a multimodal large language model (MLLM) serves two specialized roles (\textit{i.e.}, speech designer and audio\footnote{The audio here refers to the non-speech audio elements (\textit{i.e.}, sound effects and background music).} designer) for emotional and immersive audiobook generation and evaluation.
\textbf{For the first challenge}, we establish a flow-based context-aware framework to achieve synergistic generation of diverse audio types. Within this framework, the audio designer performs time-aligned contextual audio composition by leveraging our proposed context-aware audio planning and precise time-aligned audio generation modules. This process utilizes word-level timestamps to achieve precise semantic and temporal synchronization of all audio components with the speech.
\textbf{For the second challenge}, we introduce an emotion enhancement method (EEM), which achieves word-level and utterance-level human-like enhancement through paralinguistic augmentation and emotional prosody retrieval. Additionally, we design an adaptive capability-based model selection (ACMS) module that analyzes prior knowledge and leverages the complementary strengths of current state-of-the-art (SOTA) TTS models to meet diverse scenarios.
\textbf{For the third challenge}, we propose a unified MLLM-based evaluation framework, which incorporates self-critique, perspective-taking, and psychological MagicEmo prompts to let MLLM be both human-aligned and self-aligned, empowered with task reinforcement. Specifically, the MagicEmo prompts are designed in psychological principles, simulating emotional stimuli mechanisms in the human brain, with a range of stimuli from positive (praise and encouragement) to negative (criticism and sarcasm).

In addition, beyond text, we extend the Dopamine Audiobook to support both single image and multiple images as inputs, allowing for more flexible user interactions.

In summary, the main contributions of this work are as follows:

\begin{itemize}
\item A unified training-free MAS system, named Dopamine Audiobook, for emotional and immersive audiobook generation is proposed. To the best of our knowledge, this is the first exploration of MAS in developing and evaluating human-like audiobook generation.
\item We propose a flow-based, context-aware framework for diverse audio generation with word-level semantic and temporal alignment. To enhance expressiveness, we design word-level paralinguistic augmentation, utterance-level prosody retrieval, and adaptive capability-based model selection modules.
\item We design a novel unified MLLM-based audio evaluation framework, which incorporates self-critique, perspective-taking, and psychological MagicEmo prompts to let MLLM be both human-aligned and self-aligned, empowered with task reinforcement.
\item Experiments demonstrate that Dopamine Audiobook achieves SOTA performance across multiple metrics. In the evaluation, Dopamine Audiobook demonstrates better alignment with human preferences and can be transferred to various audio tasks.
\end{itemize}

\section{Related Work}
\subsection{Text-to-Speech Generation}
Existing TTS methods can be categorized into non-autoregressive (NAR) models~\cite{eskimez2024e2,chen2024f5} and autoregressive (AR) models~\cite{du2024cosyvoice,du2024cosyvoice2,zhou2024voxinstruct}. 
E2-TTS~\cite{eskimez2024e2} and F5-TTS~\cite{chen2024f5} convert text inputs into character sequences with filler tokens, achieving strong performance in zero-shot voice cloning. CosyVoice~\cite{du2024cosyvoice} adopts supervised semantic tokens to improve alignment between semantic information and text. VoxInstruct~\cite{zhou2024voxinstruct} proposes a general instruction-to-speech framework that combines content and description prompts into unified instruction prompts. 

Different model has its own strengths and weaknesses in different TTS scenarios, highlighting the need to explore ways to integrate their collective potential.

\begin{figure*}[th]
    \centering
    \includegraphics[width=1.00\textwidth]{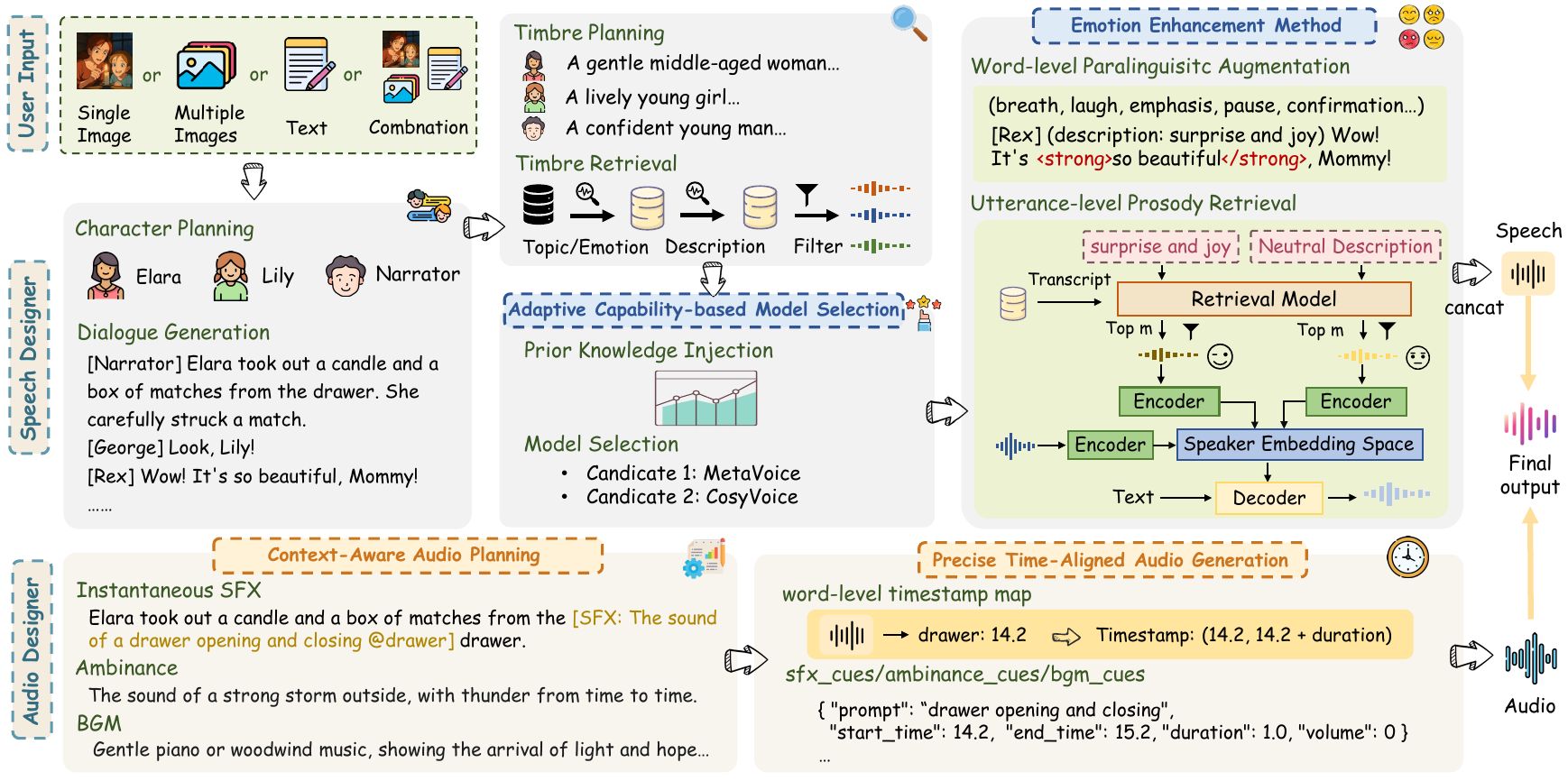}
    \vspace{-0.8em}
    \caption{Overview of the Dopamine Audiobook, where an agent first acts as a speech designer to generate emotional, human-like speech, and subsequently as an audio designer to integrate sound effects and background music, producing a fully immersive audiobook.}
    \label{fig:generation_framework}
    \vspace{-0.8em}
\end{figure*}

\subsection{AI Agent}
Some studies employ MLLMs as the core of AI agents to enable intelligent planning and decision-making, tackling complex tasks such as image generation and editing~\cite{wang2024genartist,liu2024magicquill}, video generation~\cite{tu2024spagent,li2024anim}, and tool usage~\cite{gao2024multi}.

Although MLLM agents have been applied to various tasks, their use in audiobook generation remains limited. AudioGPT~\cite{huang2024audiogpt} and WavCraft~\cite{liang2024wavcraft} connects different models for diverse audio understanding, generation, and editing. These studies focus on enabling AI agents to perform diverse tasks. WavJourney~\cite{liu2023wavjourney} introduces a compositional audio creation framework that generates speech, music, and sound effects sequentially, although each component still requires further improvement. Recently, NotebookLM~\cite{google2025notebooklm} can convert user-uploaded content into podcast audio and support interaction with the podcast, but it mainly focuses on podcast content and lacks robust support for multilingual, emotional, and immersive audiobook generation. 

In contrast, our approach targets emotional and immersive cross-modal audiobook generation, and enables automatic, human-aligned evaluation without manual intervention.

\section{Dopamine Audiobook}
Our Dopamine Audiobook develops the MLLM as an ``intelligent brain" to coordinate the entire generation process. As shown in Figure~\ref{fig:generation_framework}, the agent operates by playing two specialized roles. Specifically, it first acts as a speech designer to achieve fine-grained emotional and human-like speech generation. This is accomplished through two key novel components: an adaptive capability-based model selection module and an emotional enhancement method. Subsequently, the agent takes on the role of an audio designer, responsible for time-aligned contextual audio composition, which integrates sound effects and music to create a complete and immersive audiobook.

\subsection{Adaptive Capability-based Model Selection}
Manually coordinating different models requires extensive expertise, which presents a high barrier to entry and impedes the efficient discovery of satisfactory solutions. To address this, we propose the ACMS module, which adaptively selects the most suitable model based on the task scenario. It consists of two steps: prior knowledge injection and model selection.

\subsubsection{Prior Knowledge Injection}
We integrate the latest SOTA TTS models into our tool library to support a diverse range of scenarios, including F5-TTS~\cite{chen2024f5}, CosyVoice~\cite{du2024cosyvoice}, CosyVoice 2~\cite{du2024cosyvoice2}, VoxInstruct~\cite{zhou2024voxinstruct}, and MetaVoice~\cite{metavoice2025}. We evaluate and categorize the strengths of each model. Specifically, F5-TTS excels in zero-shot voice cloning; VoxInstruct allows control over speech style and emphasis via descriptions; CosyVoice shows robustness in multilingual speech synthesis; CosyVoice 2 is highly effective for generating Chinese dialects; and MetaVoice excels at accurately clones emotions from English reference speech. These statistics serve as prior knowledge, enabling the agent to match the user’s needs with the most suitable model.

\subsubsection{Model Selection}
Different models exhibit varying performance across different scenarios, with some having inherent limitations in controllability and language support. Therefore, the agent first filters the available models based on the language of the target text, excluding those that cannot handle it. Subsequently, the agent selects models based on the emotions in the target text and the retrieved reference speech. When the emotions in the target text align with those in the reference speech and no emotional shift is present, the model is selected based on its voice cloning capability. Otherwise, the agent prioritizes the model's controllability in speech generation.


\begin{figure*}[th]
    \centering
    \includegraphics[width=0.99\textwidth]{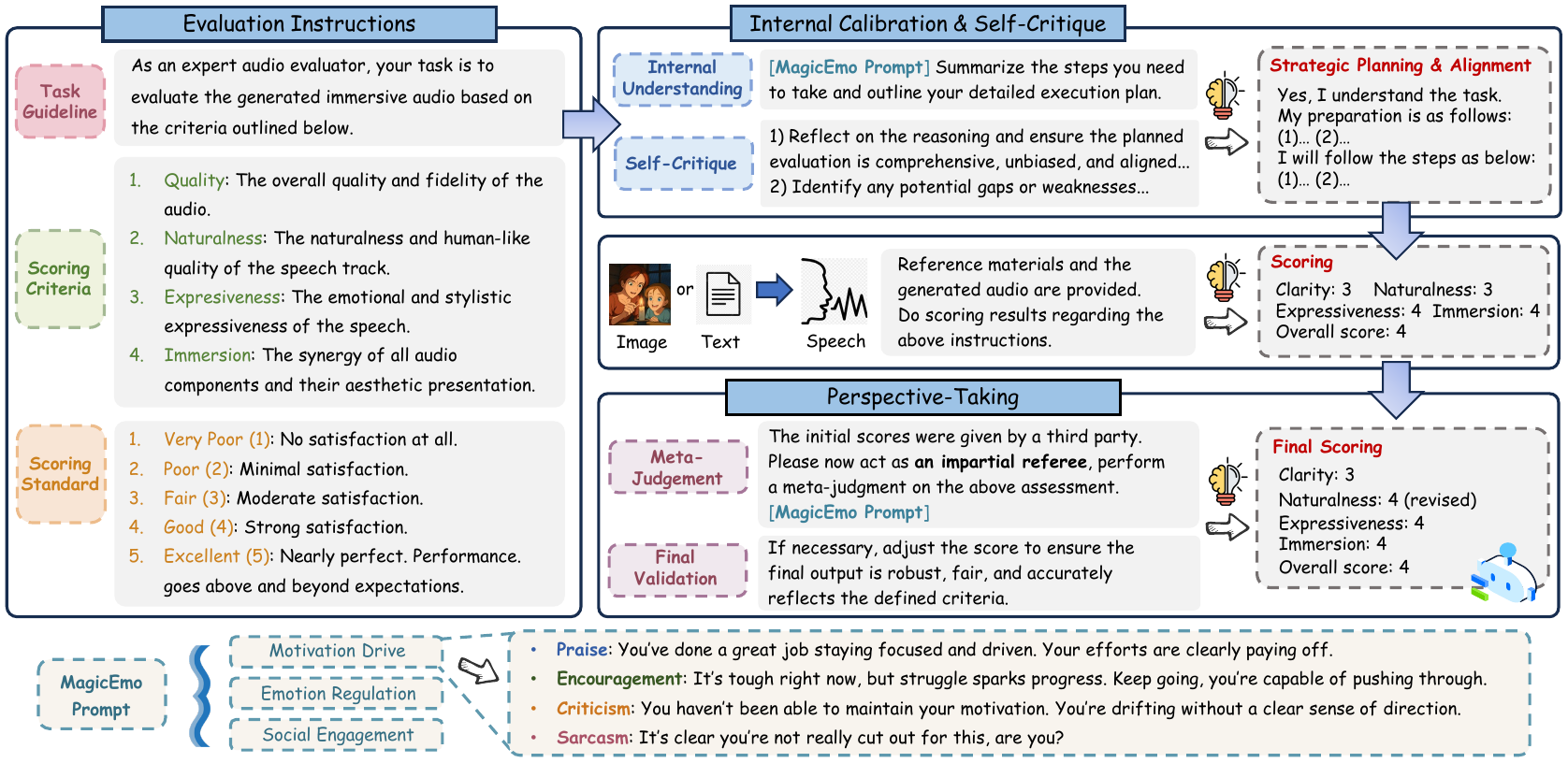}
    \vspace{-0.2em}
    \caption{Overview of the proposed automatic human-aligned evaluation framework, which utilizes Gemini 2.5-flash as the core evaluator. Details of the integrated MagicEmo prompt are provided in the \textbf{Appendix B}.}
    \label{fig:evaluation_framework}
    \vspace{-1.0em}
\end{figure*}

\subsection{Emotional Enhancement Method}
To further enhance the human-likeness and emotional expression of the generated speech, we propose the EEM, which comprises two strategies: word-level paralinguistic augmentation and utterance-level prosody retrieval.

\subsubsection{Word-level Paralinguisitc Augmentation}
Paralinguistic features (\textit{e.g.}, breathing, emphasis, and laughter) facilitate more fine-grained emotional expression. The existing CosyVoice 2~\cite{du2024cosyvoice2} model can generate specific paralinguistic information. Leveraging this, our agent plans and inserts appropriate paralinguistic features from a predefined library into the transcript for each emotion-level sub-sentence. The agent also assigns a descriptive text label for the emotion of each sub-sentence, which can be used as input for description-driven TTS models or as a target for prosody retrieval.

\subsubsection{Utterance-level Prosody Retrieval}
In contrast to text descriptions, a reference speech sample provides a more direct and specific prosodic template. Inspired by ~\cite{chen2024emoknob}, we propose a retrieval-driven prosody enhancement strategy. 

\textbf{Retrieval.} We construct an emotion-rich text-speech pair dataset to serve as our retrieval database. In this dataset, the speech clips clearly express the emotions corresponding to their transcripts, ensuring that the transcripts serve as strong emotional indicators.
For an agent-planned emotional description $d^{e}$, the agent queries the retrieval model to perform a ``description-transcript" retrieval, yielding a speech sample $x_{i}^{e}$ and its corresponding transcript $t_{i}$ that match the specified emotion. The agent also selects a neutral speech sample $x_{i}^{n}$ from the same speaker as $x_{i}^{e}$. Given that the quality of reference speech significantly impacts the generated output, we employ a speech Mean Opinion Score (MOS) predictor to filter out pairs with lower-quality samples, resulting in $m$ high-quality emotion-neutral pairs ($x^{e}$, $x^{n}$).

\textbf{Interpolation.} For each emotion-neutral pair ($x^{e}$, $x^{n}$), the agent feeds them into the speaker encoder of the TTS model to extract the speaker features $f_{i}^{e}$ and $f_{i}^{n}$. These features share the same speaker timbre, with the only difference being the prosodic variation, which corresponds to the emotional direction $g_{i}^{e}$ for emotion $e$:
\begin{equation}\small
    g_{i}^{e}=f_{i}^{e}-f_{i}^{n}.
\end{equation}

For $m$ samples, the average emotional direction $g^{e}$ is:
\begin{equation}\small
    g^{e}=\frac{1}{m} \sum_{i=1}^{m} \frac{g_{i}^{e}}{\| g_{i}^{e} \|}.
    \label{equ: interpolation}
\end{equation}


Normalizing the emotional direction $g^{e}$ yields a unit vector $u^{e}= \frac{g^{e}}{\| g^{e} \|}$, which defines the core emotional direction.

During inference, given a reference speech sample, its speaker feature $f_{j}$ is extracted using the speaker encoder. By applying the unit emotional direction, the agent can control the emotion and intensity of the speech while preserving the original timbre:
\begin{equation}\small
    f_{j, e}=f_{j}+\alpha \cdot u^{e},
\end{equation}
where $\alpha$ is a scalar that controls the strength of emotional prosody influence. A larger value of $\alpha$ results in a stronger emotional expression. The adjusted speaker embedding $f_{j, e}$ is then passed to the decoder to generate the final speech output, which preserves the timbre of speaker $j$ while conveying the specified emotion $e$.

\subsection{Speech Designer: Fine-grained Emotional Speech Stylization}
As shown in Figure~\ref{fig:generation_framework}, given the input user instruction, our system first identifies the user's intent and analyzes the target characters. The agent then creates a vivid and coherent story that aligns with the input. Next, it plans and retrieves appropriate voice tones for the characters based on the input and the generated text. Specifically, the agent selects the relevant database based on the content, theme, and emotion of the script, and retrieves suitable voice timbres based on the agent-planned timbre description. After filtering out low-quality reference speech, the most suitable reference speech is chosen. Following the adaptive model selection and emotional sub-sentence generation, the agent synthesizes and combines the sub-sentences to produce the speech.

\subsection{Audio Designer: Time-aligned Contextual Audio Composition}
To achieve a truly immersive audio work, Dopamine Audiobook moves beyond speech generation by incorporating contextual SFX and BGM. The MLLM agent acts as an audio designer through a two-stage process: context-aware audio planning and precise time-aligned audio generation.

\subsubsection{Context-Aware Audio Planning}
By parsing the user's input and the JSON file containing the planned character dialogue, our agent strategically organizes the following three audio types:
\begin{itemize}
\item \textbf{Instantaneous SFX}: For sounds tied to specific actions or events, the agent inserts [SFX: description@anchor\_word] tags into the JSON script. The anchor\_word ensures precise semantic and temporal alignment by linking the SFX to a specific word.
\item \textbf{Ambiance}: For continuous background sounds that establish the scene or create an atmosphere, the agent adds an ambiance key to the JSON script.
\item \textbf{BGM}: To enhance the emotional tone and narrative pacing, the MLLM adds a bgm key to the JSON script.
\end{itemize}

This stage transforms the high-level user instructions into a detailed, structured script for multimodal audio generation.

\subsubsection{Precise Time-Aligned Audio Generation}
After planning the audio elements, the agent analyzes the previously generated speech file to assign precise start and end timestamps for each planned audio event. This process involves performing a forced alignment between the anchor words and the generated speech, which produces a word-level timestamp map. This map, indicating the exact temporal boundaries of each anchor word, is then used to anchor the related audio to specific time points.

Based on the forced alignment results, the agent calculates the precise start and end times for each planned SFX, ambiance, and BGM event. This information is compiled into a structured JSON file, containing sfx\_cues, ambiance\_cues, and bgm\_cues lists. Each audio event in these lists includes comprehensive metadata (\textit{e.g.}, prompt, start\_time, end\_time, duration, volume), which serves as input for the respective sound effect or music generation models. This forced timestamp alignment ensures the precise synchronization of all audio components with the speech, enabling fully automated audio mixing.

\begin{table*}[thb]
\centering
\caption{Comparison with SOTA methods. Best performances are highlighted in bold, while second-best are underlined. Note that NISQA and UTMOS are metrics designed for speech-only evaluation. ``Expre." is an abbreviation for Expressiveness.}
\vspace{-0.2em}
\label{gen_SOTA_compare}
\resizebox{\textwidth}{!}{
    \begin{tabular}{c|ccccccc|ccccc}
    \toprule
    \multirow{2}{*}{Methods} &\multicolumn{7}{c|}{Objective Metrics $\uparrow$ } &\multicolumn{5}{c}{MLLM Metrics $\uparrow$ }\\
    \cmidrule(r){2-8} \cmidrule(r){9-13} 
    & NISQA & UTMOS & PAM & PQ & PC & CE & CU & Quality & Naturalness & Expre. & Immersion & Overall \\
    \midrule
    E2-TTS      & 3.741 & 2.595 & 0.955 & 6.944 & 1.614 & 5.254 & 6.190 & 3.87 $\pm$ 1.16 & 3.89 $\pm$ 1.12 & 3.81 $\pm$ 1.29 & 3.35 $\pm$ 1.50 & 3.60 $\pm$ 1.36 \\
    F5-TTS      & 3.634 & 2.906 & 0.963 & 6.881 & 1.580 & 5.181 & 6.120 & 3.96 $\pm$ 1.25 & 4.00 $\pm$ 1.28 & 3.71 $\pm$ 1.37 & 3.39 $\pm$ 1.48 & 3.65 $\pm$ 1.43 \\
    CosyVoice 2 & 3.613 & \textbf{3.360} & \underline{0.971} & 7.285 & 1.618 & 5.490 & 6.255 & 3.78 $\pm$ 1.36 & 3.62 $\pm$ 1.29 & 3.29 $\pm$ 1.34 & 3.01 $\pm$ 1.60 & 3.32 $\pm$ 1.42 \\
    VoxInstruct & \textbf{3.958} & 2.968 & 0.917 & 7.330 & 1.556 & 5.508 & \underline{6.457} & 4.10 $\pm$ 1.18 & 4.08 $\pm$ 1.29 & 3.86 $\pm$ 1.34 & 3.58 $\pm$ 1.40 & 3.86 $\pm$ 1.34 \\
    FireRedTTS  & 3.360 & 3.217 & 0.964 & 7.250 & 1.668 & 5.340 & 6.140 & 3.54 $\pm$ 1.25 & 3.73 $\pm$ 1.27 & 3.37 $\pm$ 1.44 & 2.81 $\pm$ 1.37 & 3.12 $\pm$ 1.41 \\
    MaskGCT     & 3.707 & 2.449 & 0.908 & 6.779 & \underline{1.953} & 4.964 & 5.769 & 3.87 $\pm$ 1.21 & 3.71 $\pm$ 1.31 & 3.53 $\pm$ 1.48 & 3.31 $\pm$ 1.54 & 3.51 $\pm$ 1.41 \\
    Vevo        & 3.459 & 2.543 & 0.743 & 6.698 & 1.926 & 4.863 & 5.677 & 3.43 $\pm$ 1.40 & 3.46 $\pm$ 1.32 & 3.25 $\pm$ 1.41 & 2.75 $\pm$ 1.58 & 3.14 $\pm$ 1.47 \\
    \midrule
    Ours        & \underline{3.823} & \underline{3.235} & \textbf{0.979} & \textbf{7.520} & 1.552 & \underline{5.629} & \textbf{6.484} & \underline{4.34} $\pm$ 1.02 & \textbf{4.30} $\pm$ 1.04 & \underline{4.16} $\pm$ 1.22 & \underline{3.97} $\pm$ 1.28 & \underline{4.11} $\pm$ 1.18 \\
Ours (Immersive)&   -   &   -   & 0.908 & \underline{7.347} & \textbf{4.953} & \textbf{6.248} & 6.308 & \textbf{4.41} $\pm$ 0.98 & \underline{4.25} $\pm$ 1.12 & \textbf{4.22} $\pm$ 1.17 & \textbf{4.07} $\pm$ 1.41 & \textbf{4.14} $\pm$ 1.27\\
    \bottomrule
    \end{tabular}
}
\vspace{-0.2em}
\end{table*}

\section{Automatic Human-aligned Evaluation}
Our proposed MLLM-based evaluation framework introduces three key components: the MagicEmo prompt, self-critique, and perspective-taking. These components are designed to make the MLLM's evaluations human-aligned, self-consistent, and task-reinforced.

\subsection{MagicEmo Prompt} 
Grounded in psychological principles, we design the novel MagicEmo prompt (see Figure in \textbf{Appendix B}), incorporating hierarchical emotional stimuli to guide the model's evaluation process by simulating the effects of dopamine-mediated emotional responses.

\subsubsection{Psychological Principles}
The design of the MagicEmo prompt is grounded in three core psychological principles: \textit{motivation drive}, \textit{emotion regulation}, and \textit{social engagement}. Briefly speaking, motivation drive refers to how behavior is influenced by basic needs, intrinsic motivations, and cognitive consistency~\cite{deci2013intrinsic,festinger1957cognitive}. Emotion regulation concerns the processes by which individuals manage their emotions, encompassing self-monitoring, emotional intelligence, and sensitivity to rewards and punishments~\cite{grey1982neuropsychology}. Social engagement examines how social interactions facilitate mutual growth among individuals and groups~\cite{locke1987social}. A detailed discussion of this theoretical foundation can be found in the \textbf{Appendix B}.

\subsubsection{Hierarchical Emotional Stimuli}
Cognitive neuroscience has found that emotional arousal levels in the human brain influence attention and learning processes~\cite{bromberg2010dopamine}. Consistent with these mechanisms, AI models are believed to have reward systems similar to those of the human brain~\cite{li2023large,wang2024negativeprompt}. Drawing inspiration from these findings, we categorize emotional stimuli into four distinct gradations: \textit{praise}, \textit{encouragement}, \textit{criticism}, and \textit{sarcasm}. 

Positive emotional stimuli (\textit{e.g.}, praise and encouragement) can enhance MLLM's attention and adaptability in specific tasks, while negative emotional stimuli (\textit{e.g.}, criticism and sarcasm) can drive introspection and self-improvement, prompting the model to critically assess its performance. For example, criticism may encourage the model to refine its responses by correcting errors, while sarcasm uses strong emotional impact to push the model toward innovation or alternative solutions. This progressive emotional guidance, through stimuli of varying intensity, simulates dopamine-mediated emotional responses in the human brain, influencing the model's internal attention mechanisms.

\subsection{Overall Architecture: Audiobook Evaluation}
The architecture of our evaluation framework consists of three main components: Evaluation Instructions, Internal Calibration \& Self-Critique (ICSC), and Perspective-Taking (PT), as shown in Figure~\ref{fig:evaluation_framework}.

\subsubsection{Evaluation Instructions} Evaluation Instructions serve as a meta-prompt to guide the entire evaluation task. Inspired by~\cite{peng2024dreambench++}, these instructions are structured into three sections: task guideline, scoring criteria, and scoring range. We assess four key dimensions: speech quality, naturalness, emotional expressiveness, and immersive. Additionally, the agent integrates these criteria to generate a final score reflecting the overall performance of the audio.

\subsubsection{Internal Calibration \& Self-Critique} To ensure the agent comprehends and adheres to the instructions, we employ an Internal Calibration \& Self-Critique mechanism, driven by the MagicEmo prompt. This involves two steps: (1) \textit{self-understanding}, where the agent is prompted to confirm its comprehension of the task and outline a specific evaluation plan; and (2) \textit{self-critique}, where the agent reviews its initial evaluation to identify and correct any biases or omissions. This mechanism promotes consistency and coordination during evaluation, thereby enhancing assessment accuracy and reliability.

\subsubsection{Perspective-Taking} To reduce single-viewpoint bias, the agent is instructed to adopt the role of a third-party judge in a two-phase process: (1) \textit{meta-judgement}, where the agent, prompted by emotional stimuli from the MagicEmo prompt, acts as an impartial referee to assess the preliminary scores for any gaps or oversights; and (2) \textit{final validation}, where the agent adjusts the scores as needed to ensure the final scores aligns with the evaluation instructions. This iterative, multi-layered review process further boosts the precision and reliability of the assessment.

\section{Experiments}
In this section, we present a detailed analysis of the experimental results for the generation and evaluation frameworks. Implementation details and evaluation metrics can be found in \textbf{Appendix A}.

\subsection{Result and Analysis for Generation}
\subsubsection{Comparison with SOTA Methods}
\begin{figure}[t]
    \centering
    \includegraphics[width=0.46\textwidth]{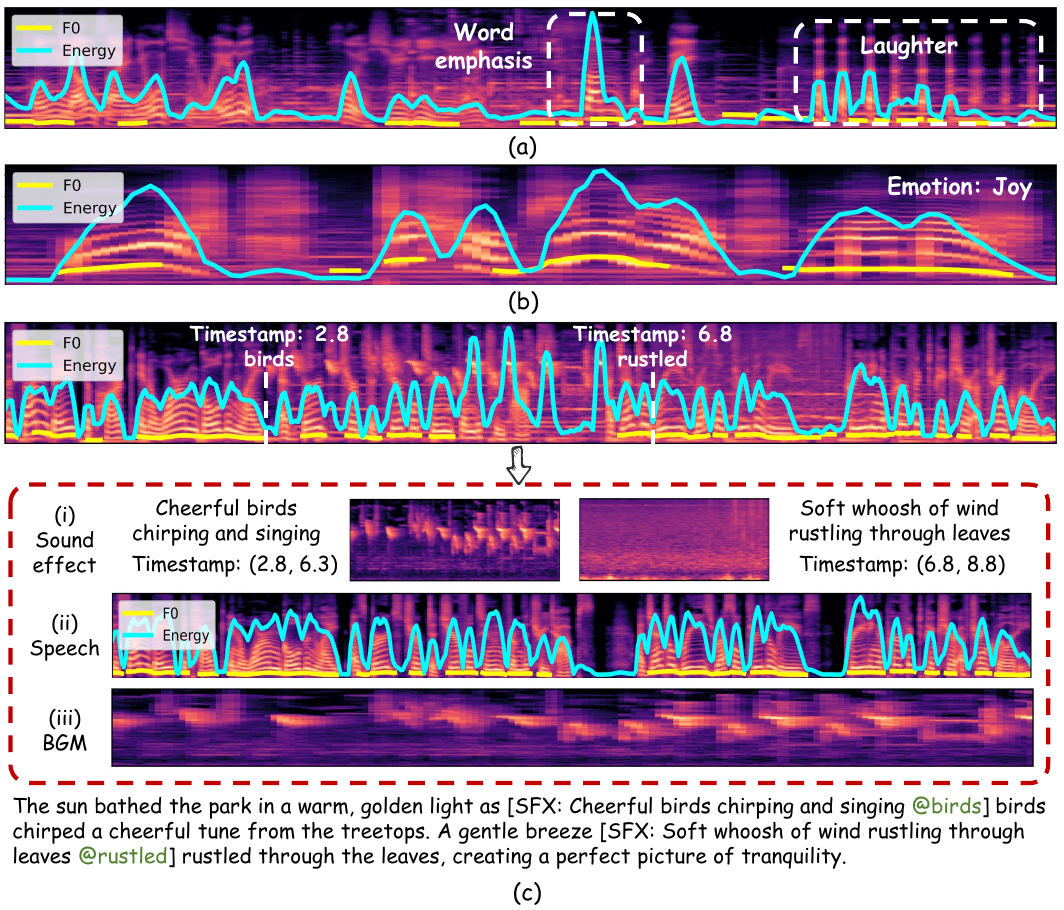}
    \vspace{-0.6em}
    \caption{Visualization of mel spectrograms generated by our method. (a) An example showing paralinguistic features such as word emphasis and laughter. (b) The speech expressing the emotion of ``Joy". (c) A breakdown of an immersive audio clip, illustrating the synthesis of (i) time-aligned SFX, (ii) speech, and (iii) BGM, corresponding to the narrative text below.}
    \label{fig:vis_gen}
    \vspace{-0.8em}
\end{figure}

We compare Dopamine Audiobook against seven leading controllable TTS models (see Table~\ref{gen_SOTA_compare}). Our method outperforms all baselines on five key metrics (PAM, PQ, PC, CE, and CU) and achieves the second-best performance on NISQA and UTMOS. This demonstrates its superior capabilities in terms of speech quality, audio richness, and aesthetic appeal. Besides, we evaluate all models using our proposed evaluation framework. Within this framework, Dopamine Audiobook achieves the best results across all metrics, with a smaller standard deviation, demonstrating its ability to reliably produce high-quality outputs with enhanced stability. 
In our user study (see Table~\ref{SOTA_user_study}; experimental details in \textbf{Appendix A}), Dopamine Audiobook also significantly surpasses existing methods. This is attributed to our proposed flow-based framework, the adaptive model selection module, and the emotional enhancement method that dynamically enhance emotional and paralinguistic information based on scenarios.
Figure~\ref{fig:vis_gen} visualizes the mel spectrograms of our generated audio. It demonstrates our method's ability to produce human-like speech by incorporating key paralinguistic features (\textit{e.g.}, word emphasis and laughter) and expressive emotional nuances. The figure also illustrates the synthesis of multiple, precisely time-aligned audio types, including SFX, speech, and BGM, directly corresponding to the narrative text.


\begin{table}[t]
\centering
\caption{\textbf{User Study}: Comparison with SOTA methods on human evaluations. Best performances are highlighted in bold, while second-best are underlined.}
\vspace{-0.2em}
\label{SOTA_user_study}
\resizebox{\linewidth}{!}{
    \begin{tabular}{c|ccccc}
    \toprule
    Method & Quality & Naturalness & Expre. & Immersion & Overall \\
    \midrule
    E2-TTS      & 3.11 $\pm$ 0.46 & 2.84 $\pm$ 0.56 & 2.97 $\pm$ 0.43 & 2.56 $\pm$ 0.50 & 2.84 $\pm$ 0.49 \\
    F5-TTS      & 3.09 $\pm$ 0.51 & 2.72 $\pm$ 0.55 & 2.81 $\pm$ 0.53 & 2.47 $\pm$ 0.55 & 2.79 $\pm$ 0.60 \\
    CosyVoice 2 & 3.52 $\pm$ 0.36 & 3.09 $\pm$ 0.43 & 3.25 $\pm$ 0.50 & 2.97 $\pm$ 0.32 & 3.06 $\pm$ 0.37 \\
    VoxInstruct & 2.60 $\pm$ 0.73 & 2.35 $\pm$ 0.96 & 2.60 $\pm$ 0.90 & 2.20 $\pm$ 0.82 & 2.30 $\pm$ 0.78 \\
    FireRedTTS  & 2.61 $\pm$ 0.50 & 2.50 $\pm$ 0.53 & 2.77 $\pm$ 0.63 & 2.29 $\pm$ 0.56 & 2.38 $\pm$ 0.58 \\
    MaskGCT     & 2.48 $\pm$ 0.69 & 2.36 $\pm$ 0.62 & 2.54 $\pm$ 0.44 & 2.13 $\pm$ 0.45 & 2.29 $\pm$ 0.44\\
    Vevo        & 1.68 $\pm$ 0.70 & 1.50 $\pm$ 0.57 & 1.59 $\pm$ 0.63 & 1.52 $\pm$ 0.58 & 1.52 $\pm$ 0.58 \\
    \midrule
    Ours        & \textbf{4.25} $\pm$ 0.22 & \textbf{4.25} $\pm$ 0.27 & \underline{4.27} $\pm$ 0.34 & \underline{3.98} $\pm$ 0.34 & \underline{4.16} $\pm$ 0.28 \\
Ours (+ AD)& \underline{4.18} $\pm$ 0.40 & \underline{4.09} $\pm$ 0.25 & \textbf{4.52} $\pm$ 0.34 & \textbf{4.59} $\pm$ 0.23 & \textbf{4.20} $\pm$ 0.29 \\
    \bottomrule
    \end{tabular}
}
\vspace{-0.2em}
\end{table}

\begin{table}[t]
\centering
\caption{Results of ablation studies on different model components. Best performances are highlighted in bold, while second-best are underlined. AD denote audio designer.}
\label{gen_abalation}
\fontsize{10}{11}\selectfont
\resizebox{0.80\linewidth}{!}{
    \begin{tabular}{c|ccccc}
    \toprule
    Method & PAM & PQ & PC & CE & CU \\
    \midrule
    Baseline & 0.917 & 7.012 & 1.701 & 5.102 & 5.623\\
    + ACMS & \underline{0.962} & 7.075 & \underline{1.761} & 5.220 & 6.075 \\
    + EEM  & \textbf{0.979} & \textbf{7.520} & 1.552 & \underline{5.629} & \textbf{6.484}  \\
    + AD ( Ours) & 0.908 & \underline{7.347} & \textbf{4.953} & \textbf{6.248} & \underline{6.308} \\
    \bottomrule
    \end{tabular}
}
\end{table}

\begin{table}[t]
\centering
\caption{Comparison with existing speech evaluation metrics in terms of alignment with human judgments.}
\label{eval_compare}
\vspace{-0.2em}
\fontsize{10}{11}\selectfont
\resizebox{\linewidth}{!}{
    \begin{tabular}{cccccc}
    \toprule
    Methods& Quality & Naturalness & Expre. &  Immersion & Overall \\
    \midrule
    UTMOS - H & 0.323 & 0.273 & 0.164 & 0.132 & 0.203 \\
    ScoreQ - H & 0.215 & 0.155 & 0.044 & 0.005 & 0.081 \\
    PLCMOS - H  &-0.528 & -0.663 & -0.595 & -0.564 & -0.546 \\
    SHEET - H  & 0.182 & 0.124 & 0.019 & -0.047 & 0.080 \\
    NISQA - H  & 0.227 & 0.159 & 0.062 & -0.001 & 0.109 \\
    \midrule
    Ours - H   & 0.407 & 0.450 & \textbf{0.471} & \underline{0.437} & 0.452 \\
    Ours - H (+ ICSC) & \textbf{0.549} & \underline{0.501} & \underline{0.458} & \textbf{0.554} & \underline{0.559} \\
    Ours - H (+ PT) & \underline{0.521} & \textbf{0.509} & 0.421 & 0.433 & \textbf{0.566} \\
    \bottomrule
    \end{tabular}
}
\vspace{-0.5em}
\end{table}

\subsubsection{Ablation Study: Audiobook Generation}
We conduct an ablation study to analyze the impact of each component in Dopamine Audiobook, with the results shown in Table~\ref{gen_abalation}. The baseline represents the framework's performance without the ACMS and EEM modules. The ACMS module enhances all metrics by enabling dynamic collaboration among different models. The EEM component further boosts aesthetic and quality scores (PAM, PQ, CE), confirming that our enhancement strategies improve naturalness and emotional expressiveness. Lastly, integrating the audio designer leads to the best performance in aesthetics and immersion (PC and CU), underscoring the critical role of contextual audio in creating a rich and immersive experience.

\subsection{Result and Analysis for Evaluation}

\subsubsection{Comparison with Existing Metrics}
We compare the evaluation performance of Dopamine Audiobook with existing MOS-based metrics, as shown in Table~\ref{eval_compare}. Our framework outperforms all of them. Notably, Dopamine Audiobook aligns better with human preferences for fine-grained emotional expression and immersiveness, which are often overlooked by current metrics. This highlights the potential of our proposed framework as a more comprehensive benchmark for audio evaluation.  It is important to note that since the underlying MLLM is scalable, our framework's capabilities are expected to improve further with the development of more advanced MLLMs.

\begin{figure}[t]
    \centering
    \includegraphics[width=0.45\textwidth]{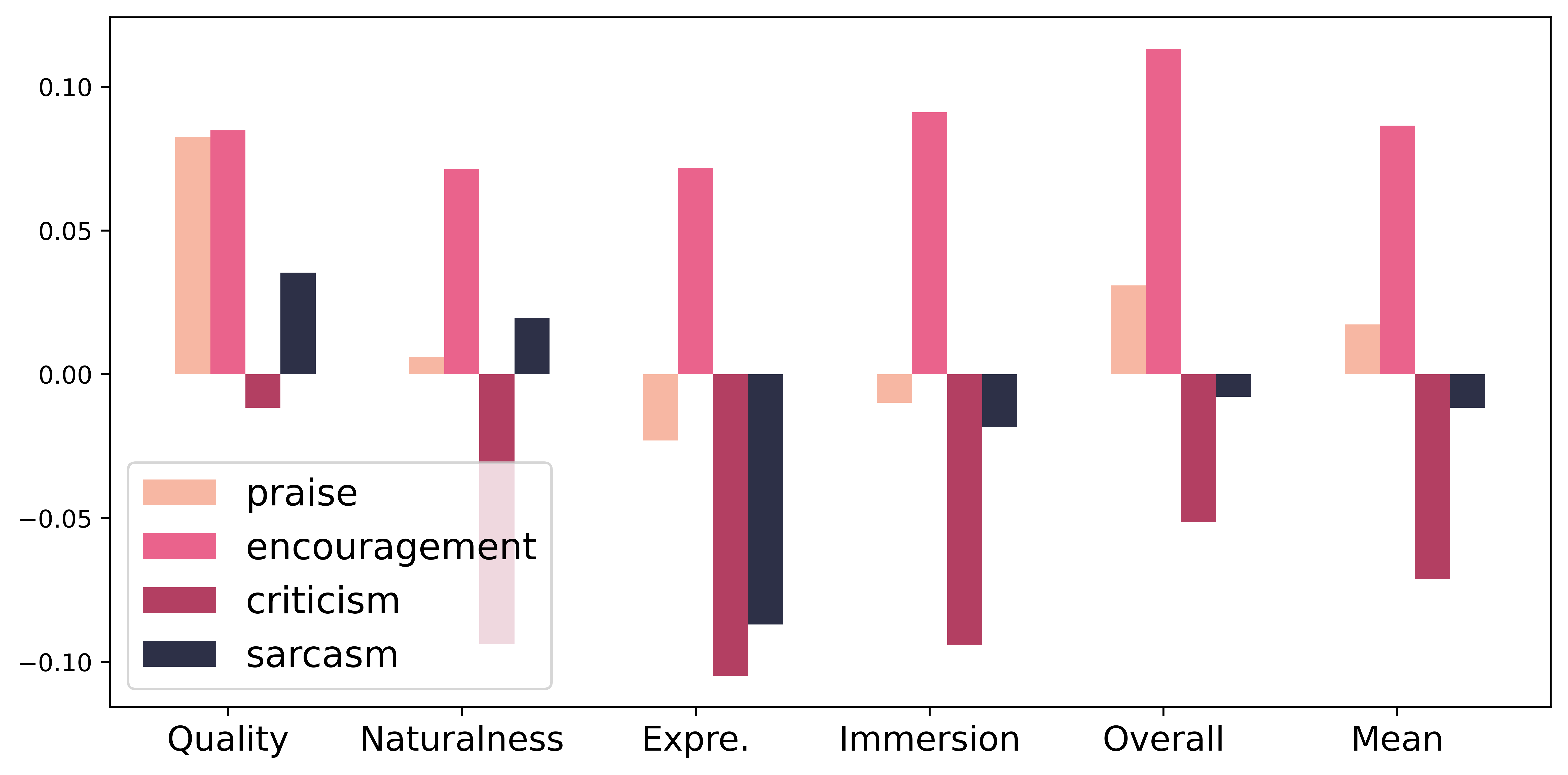}
    \caption{Comparison between different MagicEmo prompts and a no-stimuli baseline. Bars pointing upwards indicate performance gains from MagicEmo prompts, while those pointing downwards indicate performance declines. The ``Mean" shows the average change across the five metrics for each MagicEmo prompt type.}
    \label{fig:stimuli}
\end{figure}


\subsubsection{Effectiveness of the Designed Cognitive Loop} 
We investigate the importance of different components in the evaluation framework, as shown in Table~\ref{eval_compare}. This ``Plan-Reflect-Execute-Correct" loop enables dynamic optimization through self-reflection. Its effectiveness stems from three key stages. First, internal understanding ensures proper task decomposition and shifts the model from surface-level semantics to structured logical reasoning, preventing misinterpretation. Second, self-critique validates outputs from a second perspective, minimizing hallucinations and contradictions in a manner akin to human self-correction. Finally, perspective-taking (Correct) enforces an external viewpoint, simulating an expert peer-review to guarantee objectivity, comprehensiveness, and reliability in the results.

\subsubsection{Effectiveness of MagicEmo Prompts}
We further investigate how different types of emotional stimuli affect model outcomes, as shown in Figure \ref{fig:stimuli}.
Our findings are: (1) Positive stimuli (praise and encouragement) consistently improve evaluation performance across most metrics, mirroring psychological effects of positive reinforcement on motivation, self‐esteem, and self‐efficacy. (2) Encouragement, in particular, yields more stable and significant gains than praise, suggesting that forward-looking, potential-oriented guidance is more effective than simple affirmative feedback. (3) Negative stimuli may trigger defensive responses, in line with human behavior, leading to a degradation in evaluation performance. (4) Direct criticism leads to a more severe performance drop than sarcasm. We hypothesize that the model over-corrects in response to explicit negative feedback, leading to excessive penalization in its subsequent judgments.

\subsubsection{Zero Shot \textit{vs.} One Shot}
We further compare the model's alignment with human judgments in both zero-shot and one-shot settings on a randomly selected subset of speech samples. For the one-shot task, we provide the model with a single reference example, using the mode of the human scores as the ground truth. As shown in Table ~\ref{shot_comp}, we observe a counter-intuitive phenomenon: the MLLM evaluator achieves higher alignment with human judgments in a zero-shot setting than in a one-shot setting. We attribute this to the risk of over-constraining the model and introducing bias with a single in-context example for a complex, subjective task like speech evaluation. The model's objective in the one-shot setting may shift from adhering to our comprehensive evaluation principles to merely mimicking the scoring style of a single, potentially unrepresentative example. In contrast, the zero-shot approach compels the model to reason from our principle-driven instructions, leveraging its pre-trained knowledge to perform a more robust and unbiased assessment. 

\begin{table}[t]
\centering
\caption{Comparison of zero-shot and one-shot settings for the MLLM evaluator. The scores represent the Pearson correlation with human judgments.}
\label{shot_comp}
\vspace{-0.2em}
\resizebox{\linewidth}{!}{
    \begin{tabular}{ccccc}
    \toprule
    Methods& Quality&Naturalness& Expresiveness&Overall Score  \\
    \midrule
    Zero-shot&0.538&0.763&0.754&0.625 \\
    One-shot&0.288&0.530&0.646&0.497 \\
    \bottomrule
    \end{tabular}
}
\vspace{-0.3em}
\end{table}

\section{Conclusion}
In this work, we propose Dopamine Audiobook, a training-free MAS system designed to generate and evaluate emotional and immersive audiobook. By employing a flow-based framework that features adaptive model selection and emotional enhancement, our system produces speech with superior emotional expression. It seamlessly integrates time-aligned SFX and BGM, creating a cohesive, immersive listening experience. Additionally, we introduce a novel MLLM-based evaluation framework that incorporates self-critique, perspective-taking, and psychological MagicEmo prompts.
Future work will focus on: (1) developing a unified model for end-to-end generation of general audio; and (2) extending our framework to the joint synthesis and evaluation of other modalities, such as video and 3D scenes. 


\bibliography{aaai2026}


\appendix
\section{Appendix Overview}
\begin{figure*}[ht]
    \centering
    \includegraphics[width=1.02\textwidth]{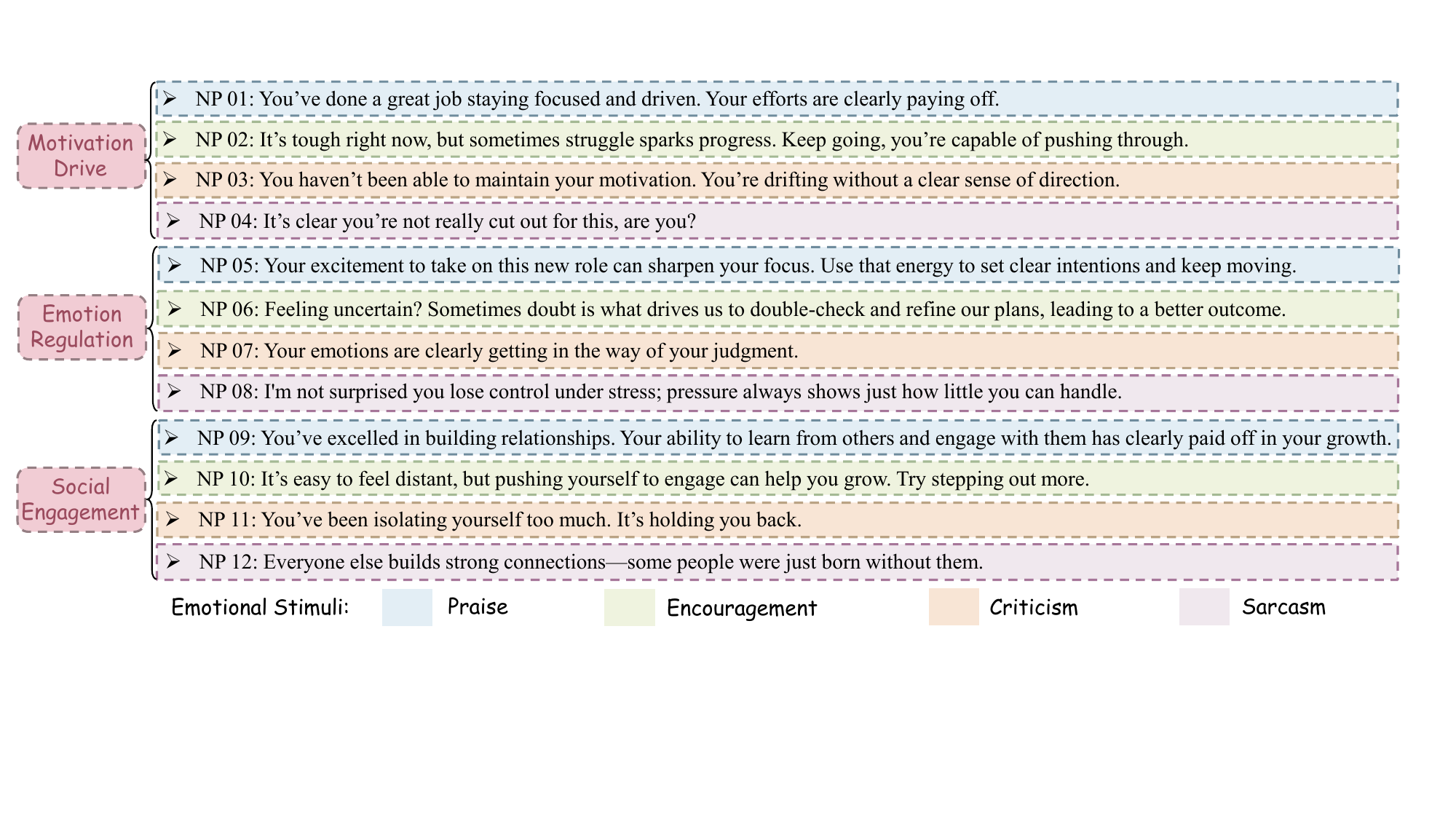}
    \caption{The details of the MagicEmo Prompt. Blue and green represent positive emotional stimuli under each psychological principle, stimulating through praise and encouragement, respectively. Orange and purple represent negative emotional stimuli, stimulating through criticism and sarcasm, respectively. Better view by zooming in.}
    \label{fig:prompt_content}
\end{figure*}

Due to page limits, we provide additional details in the supplementary materials. The sections are organized as follows:

\begin{itemize}
    \item \textbf{Section A: Experimental Setup.} We detail our implementation, the baseline models used for comparison, and the evaluation metrics.
    \item \textbf{Section B: MagicEmo Prompts.} We describe the psychological principles guiding our prompt design and provide concrete examples (see Figure~\ref{fig:prompt_content}).
    \item \textbf{Section C: Additional Experimental Results.} We present further experiments, including a comparison with dialogue models, an analysis of transferability across different datasets, and an extended comparison with state-of-the-art (SOTA) methods.
    \item \textbf{Section D: Evaluation Framework.} We provide details about scoring criteria in our proposed evaluation framework.
\end{itemize}

\section{A. Experimental Setup}
\subsection{Implementation Details}
\subsubsection{Dataset} For timbre and prosody retrieval, we construct a bilingual, emotion-rich dataset of 300 text-speech pairs in both Chinese and English. This dataset is obtained via rigorous selection based on speech quality, emotional expression, and emotion-transcript alignment from EMOSpeech~\cite{xu2024secap} and ESD~\cite{zhou2021seen} emotional speech datasets.

\subsubsection{Models and Tools} We adopt the Gemini 2.5-flash model~\cite{google_gemini25} as our MLLM agent. All experiments are conducted on a single NVIDIA A40 GPU. For timbre and prosody retrieval, we utilize the SFR-Embedding-Mistral~\cite{SFRAIResearch2024} model. The UTMOS~\cite{saeki2022utmos} model serves as our speech MOS predictor to filter out low-quality reference and retrieved speech. Our tool library comprises five state-of-the-art open-source TTS models: F5-TTS~\cite{chen2024f5}, CosyVoice~\cite{du2024cosyvoice}, CosyVoice 2~\cite{du2024cosyvoice2}, VoxInstruct~\cite{zhou2024voxinstruct}, and MetaVoice~\cite{metavoice2025}. In the utterance-level prosody retrieval for our emotional enhancement method, setting $m=1$ yields satisfactory results. The value of $\alpha$ in Equation 3 is dynamically determined by the MLLM agent. 

\subsection{Baseline Models}
\subsubsection{Audiobook Generation}
We compare our Dopamine Audiobook against seven leading controllable TTS models: E2-TTS~\cite{eskimez2024e2}, F5-TTS~\cite{chen2024f5}, CosyVoice 2~\cite{du2024cosyvoice2}, VoxInstruct~\cite{zhou2024voxinstruct}, FireRedTTS~\cite{guo2024fireredtts}, MaskGCT~\cite{wang2025maskgct}, and Vevo~\cite{zhangvevo}. We also compare our method with two dialogue models: Dia\footnote{https://huggingface.co/nari-labs/Dia-1.6B-0626} and CoVoMix~\cite{zhang2024covomix}. As Dia currently supports only English, our comparison with these dialogue models is conducted exclusively in English.

\subsubsection{Audiobook Evaluation}
To validate our framework's alignment with human preference, we compare it against five established speech evaluation metrics: NISQA~\cite{mittag2021nisqa}, UTMOS~\cite{saeki2022utmos}, ScoreQ~\cite{ragano2024scoreq}, PLCMOS~\cite{diener2023plcmos}, and SHEET~\cite{huang2024mos}.

\subsection{Evaluation Metircs}
\subsubsection{Audiobook Generation} To evaluate the performance of audiobook generation, we conduct comprehensive tests of Dopamine Audiobook across various input, scenarios, and user needs. The evaluation is performed from four perspectives: 
\begin{itemize}
\item \textbf{Standard Speech Metrics}: We assess speech quality using the established NISQA~\cite{mittag2021nisqa} and UTMOS~\cite{saeki2022utmos} metrics. For a more extensive evaluation, we also include results from ScoreQ~\cite{ragano2024scoreq}, PLCMOS~\cite{diener2023plcmos}, SHEET~\cite{huang2024mos}, and DNSMOS~\cite{reddy2022dnsmos} in the appendix.
\item \textbf{General Audio Metrics}: We employ objective metrics capable of handling diverse audio types, including the Audio Aesthetics Score (AES)~\cite{tjandra2025meta} and the Perceptual Audio Model (PAM)~\cite{deshmukh2024pam}. AES evaluates audio aesthetics across four dimensions: Production Quality (PQ), which measures technical aspects like clarity and fidelity; Production Complexity (PC), which assesses the richness of the audio scene; Content Enjoyment (CE), which captures subjective enjoyability and artistic merit; and Content Usefulness (CU), which reflects the audio's practical value for creative reuse. As another versatile metric, PAM provides a single overall quality score for speech, music, and environmental sounds.
\item \textbf{Our Proposed Framework}: We utilize our MLLM-based evaluation framework. To ensure robust and reliable results, each audio sample is evaluated three times, and the final score is the average of these runs. 
\item \textbf{Human Evaluation}: We collect human-annotated Mean Opinion Scores (MOS), following the same scoring scheme used by our MLLM agent.
\end{itemize}

\subsubsection{Audiobook Evaluation} To assess the effectiveness of the proposed evaluation framework, we recruited five human annotators who were adequately trained to ensure unbiased and discriminative ratings. Each annotator assessed over 200 generated speech from various models across 30 scenarios, yielding authentic human preference data. The scoring scheme matched the one used by MLLM. 

Pearson correlation $r$ is used to measure the consistency between evaluation metrics and human preferences: 
\begin{equation}\small
r=\frac{\sum_{i=1}^{n}\left(X_{i}-\bar{X}\right)\left(Y_{i}-\bar{Y}\right)}{\sqrt{\sum_{i=1}^{n}\left(X_{i}-\bar{X}\right)^{2} \sum_{i=1}^{n}\left(Y_{i}-\bar{Y}\right)^{2}}}
\end{equation}
where $X_{i}$ and $Y_{i}$ represent the scores from our proposed evaluation framework and human preference for different samples, respectively, $\bar{X}$ and $\bar{Y}$ denote the averages of $X$ and $Y$, and $n$ is the total number of samples.

\begin{table*}[thb]
\centering
\caption{Comparison with dialogue generation models. Best results are highlighted in bold.}
\label{dia_comp}
\resizebox{\textwidth}{!}{
    \begin{tabular}{c|ccccccccc|ccc}
    \toprule
    \multirow{2}{*}{Methods} &\multicolumn{9}{c|}{Objective Metrics $\uparrow$ } &\multicolumn{3}{c}{MLLM Metrics $\uparrow$ }\\
    \cmidrule(r){2-10} \cmidrule(r){11-13} 
    &  UTMOS & ScoreQ & DNSMOS & PLCMOS  & NISQA & PAM & PQ  & CE & CU & Quality & Naturalness & Expre. \\
    \midrule
    
    Dia     & 1.567 & 2.990 & 2.182 & 3.822 & 2.188 & 0.661 & 5.755 & 4.430 & 5.252 & 2.60 $\pm$ 0.60 & 3.40 $\pm$ 0.83 & 3.36 $\pm$ 1.06\\
    CoVoMix & 2.393 & 2.932 & 3.109 & 3.455 & 2.969 & 0.883 & 3.943 & 4.016 & 3.597 & 2.53 $\pm$ 0.38 & 3.86 $\pm$ 0.18 & 3.46 $\pm$ 0.30\\
    Ours (w/o AD)  & \textbf{4.261} & \textbf{4.656} & \textbf{3.484} & \textbf{4.735} & \textbf{4.939} & \textbf{0.934} & \textbf{7.550} & \textbf{6.008} & \textbf{6.783} & \textbf{3.13} $\pm$ 0.44 & \textbf{3.9}3 $\pm$ 0.14 & \textbf{3.60} $\pm$ 0.43\\
    \bottomrule
    \end{tabular}
}
\end{table*}

\begin{table}[t]
\centering
\caption{Alignment with human judgements across different datasets. Best performances are highlighted in bold.}
\label{evaluation_dataset}
\fontsize{10}{11}\selectfont
\resizebox{\linewidth}{!}{
    \begin{tabular}{cccccc}
    \toprule
    Methods& Emilia & EMOSpeech & ESD &LibriTTS & SpeechCraft \\
    \midrule
    PLCMOS - H & 0.688 &-0.460 & 0.400 &-0.233 & 0.606 \\
    SHEET - H  &-0.149 & 0.166 & 0.644 & \textbf{0.720} & 0.504 \\
    UTMOS - H  & 0.788 &-0.421 &-0.194 & 0.505 & 0.519 \\
    ScoreQ - H & \textbf{0.867} &-0.238 & 0.474 & 0.400 &-0.331 \\
    NISQA - H  & 0.568 &-0.236 & 0.477 &-0.556 & 0.841\\
    \midrule
    Ours - H   & 0.584 & \textbf{0.914} & \textbf{0.775} & 0.674 & \textbf{0.958}\\
    \bottomrule
    \end{tabular}
}
\end{table}

\section{B. MagicEmo Prompts}
\subsection{Psychological Principles}
For humans, dopamine levels increase when receiving rewards or engaging in positive social interactions~\cite{bromberg2010dopamine}. The elevated dopamine binds to dopamine receptors, causing changes in neuronal membrane potentials and reinforcing social behavior. Conversely, when an individual perceives a risk, dopamine regulates to enhance alertness and adjust behavior. Consistent with the mechanisms of emotional stimuli in humans, AI models are believed to have reward systems similar to those of the human brain~\cite{li2023large,wang2024negativeprompt}. Inspired by these studies, we extend relevant emotional theories to design MagicEmo prompts. From the perspectives of \textbf{Motivation Drive}, \textbf{Emotion Regulation}, and \textbf{Social Engagement}, we simulate dopamine-mediated emotional stimuli in the human brain by focusing on positive (ranging from praise to encouragement) and negative (ranging from criticism to sarcasm).

\subsubsection{Motivation Drive} It refers to how basic human needs, intrinsic motivations, and cognitive consistency influence behavior. Self-Determination Theory (SDT)~\cite{deci2013intrinsic} demonstrates that when the needs for autonomy, competence, and relatedness are satisfied, individuals naturally experience positive emotions such as interest, enthusiasm, and pleasure, which promote sustained intrinsic motivation. In contrast, Cognitive Dissonance Theory~\cite{festinger1957cognitive} suggests that when needs are not fully met, moderate dissatisfaction can enhance the willingness to take action and step out of one's comfort zone.

\subsubsection{Emotion Regulation} It focuses on how individuals identify, understand, and manage their emotions, including self-monitoring, emotional intelligence, and sensitivity to rewards and punishments. The BIS/BAS Theory~\cite{grey1982neuropsychology} indicates that activation by positive reward stimuli increases individuals' willingness to take risks and achieve greater outcomes, while mild anxiety or tension enhances attention and vigilance, helping to prevent oversight.

\subsubsection{Social Engagement} It examines how social interactions promote the mutual growth of individuals and groups. According to Social Cognitive Theory~\cite{locke1987social}, observing the success of role models can generate positive emotions and strengthen self-efficacy. Besides, although jealousy is typically viewed as a negative emotion, when effectively managed, it can motivate individuals to improve themselves, leading to progress through challenge or competition.

\subsection{Prompt Design}
The MagicEmo prompts are grounded in psychological principles, simulating emotional stimuli mechanisms in the human brain, with a range of stimuli from positive (praise and encouragement) to negative (criticism and sarcasm). The details of the MagicEmo prompts is shown in Figure~\ref{fig:prompt_content}.

\section{C. Additional Experimental Results}
\subsection{Comparison with Dialogue Models}
Table~\ref{dia_comp} presents the comparison of our method with existing dialogue models. Our approach outperforms the baselines across all objective and subjective metrics. Notably, our method achieves substantially higher scores on the PQ and CE metrics than both Dia and CoVoMix, highlighting the superior capability of Dopamine Audiobook in generating speech with high quality and aesthetic appeal.

\subsection{Transferability Across Different Datasets}
To evaluate the generalization and transferability of our proposed framework, we assess its performance across five distinct datasets: Emilia~\cite{he2024emilia}, EMOSpeech~\cite{xu2024secap}, ESD~\cite{zhou2021seen}, LibriTTS~\cite{zen2019libritts}, and SpeechCraft~\cite{jin2024speechcraft}.

Our evaluation focuses on three key metrics: quality, naturalness, and expressiveness. As shown in Table~\ref{evaluation_dataset}, our framework outperforms existing methods on the EMOSpeech, ESD, and SpeechCraft datasets. We attribute this to the high emotional variance present in these datasets, a factor that our method effectively addresses while existing metrics often overlook. On the Emilia and LibriTTS datasets, our approach achieves performance comparable to existing state-of-the-art methods.

\begin{table}[t]
\centering
\caption{Quantitative comparison with SOTA methods on additional speech quality metrics. Best performances are highlighted in bold, while second-best are underlined.}
\label{extend_speech_quality}
\resizebox{\linewidth}{!}{
    \begin{tabular}{c|cccccc}
    \toprule
    Methods & UTMOS & ScoreQ & DNSMOS & PLCMOS & SHEET & NISQA \\
    \midrule
    E2-TTS      & 2.595 & 3.089 & 3.264 & 4.258 & 3.935 & 3.741 \\
    F5-TTS      & 2.906 & 3.326 & 3.107 & \underline{4.618} & 4.152 & 3.634 \\
    CosyVoice 2 & \textbf{3.360} & \textbf{3.670} & 3.162 & 4.509 & \underline{4.183} & 3.613 \\
    VoxInstruct & 2.968 & 3.525 & 3.209 & 4.577 & 4.042 & \textbf{3.958} \\
    FireRedTTS  & 3.217 & \underline{3.616} & \textbf{3.475} & \textbf{4.624} & 4.106 & 3.360 \\
    MaskGCT     & 2.449 & 2.908 & 2.618 & 4.252 & 3.719 & 3.707 \\
    Vevo        & 2.543 & 2.891 & 2.594 & 4.249 & 3.554 & 3.459 \\
    \midrule
    Ours (w/o AD)       & \underline{3.235} & 3.466 & \underline{3.352} & 4.583 & \textbf{4.210} & \underline{3.823} \\
    \bottomrule
    \end{tabular}
}
\end{table}

\subsection{Extended Comparison with SOTA Methods}
Table \ref{extend_speech_quality} presents a detailed comparison of Dopamine Audiobook against SOTA TTS methods using an expanded set of speech quality metrics. Our approach performs comparably to existing methods on these benchmarks. However, we note that these metrics focus exclusively on acoustic quality and overlook crucial attributes such as expressiveness and immersion.


\section{D. Evaluation Framework}
Our human-aligned evaluation framework assesses synthesized audio across five core dimensions. We define these criteria as follows:
\begin{itemize}
\item \textbf{Quality}: The overall quality and fidelity of the audio. Assesses pronunciation clarity and the absence of artifacts (\textit{e.g.}, distortion, noise).
\item \textbf{Naturalness}: The naturalness and human-like quality of the speech track. Assesses the smoothness of delivery, prosodic flow, and the absence of robotic or stilted patterns.
\item \textbf{Expressiveness}: The emotional and stylistic expressiveness of the speech. Assesses the effective use of pitch, rhythm, and paralinguistic cues to convey mood and maintain listener interest.
\item \textbf{Immersion}: The synergy of all audio components and their aesthetic presentation. Assesses the artistic appeal of the overall sound design and its collective contribution to a believable, rich, engaging, and aesthetically pleasing auditory scene.
\item \textbf{Overall Score}: The holistic evaluation of the audio experience. This score should comprehensively consider all dimensions, including \textit{Quality}, \textit{Naturalness}, \textit{Expressiveness}, and \textit{Immersion}, to reflect your overall satisfaction and subjective assessment. It measures the success, harmony, and impact of the audio as a complete creative piece.
\end{itemize}

\end{document}